# Momentum-space imaging of σ-orbitals for chemical analysis

Anja Haags[1,2,3], Xiaosheng Yang[1,2,3], Larissa Egger[4], Dominik Brandstetter[4], Hans Kirschner[5], François C. Bocquet[1,2], Georg Koller[4], Alexander Gottwald[5], Mathias Richter[5], J. Michael Gottfried[6], Michael G. Ramsey[4], Peter Puschnig[4]*, Serguei Soubatch[1,2]*, F. Stefan Tautz[1,2,3]

[1]Peter Grünberg Institut (PGI-3), Forschungszentrum Jülich; Jülich, Germany.

[2]Jülich Aachen Research Alliance (JARA), Fundamentals of Future Information Technology; Jülich, Germany.

[3]Experimentalphysik IV A, RWTH Aachen University; Aachen, Germany.

[4]Institut für Physik, Karl-Franzens-Universität Graz, NAWI Graz; Graz, Austria.

[5]Physikalisch-Technische Bundesanstalt (PTB); Berlin, Germany.

[6]Fachbereich Chemie, Philipps-Universität Marburg; Marburg, Germany.

*Corresponding author. Email: peter.puschnig@uni-graz.at, s.subach@fz-juelich.de

**Tracing the modifications of molecules in surface chemical reactions benefits from the possibility to image their orbitals. While delocalized frontier orbitals with π-character are imaged routinely with photoemission orbital tomography, they are not always sensitive to local chemical modifications, particularly the making and breaking of bonds at the molecular periphery. For such bonds, σ-orbitals would be far more revealing. Here, we show that these orbitals can indeed be imaged in a remarkably broad energy range, and that the plane wave approximation, an important ingredient of photoemission orbital tomography, is also well fulfilled for these orbitals. This makes photoemission orbital tomography a unique tool for the detailed analysis of surface chemical reactions. We demonstrate this by identifying the reaction product of a dehalogenation and cyclodehydrogenation reaction.**

In frontier orbital theory, orbital patterns are regarded as important for chemical empiricism – they determine molecular reactivities and reaction pathways (*1,2*). Indeed, orbitals are attributed a reality of their own, beside the total electron densities in which influential electronic structure theories are framed (*3,4*). This has spurred many successful efforts to image orbitals (*5–7*), in spite of subtleties of the single-electron orbital concept (*8,9*). Orbital imaging has also been applied to the investigation of chemical bonding (*10–12*). The imaging of orbitals with low-temperature scanning tunneling microscopy (STM) (*6,10,11*) is particularly attractive, because it combines spectroscopic energy resolution with sub-angström spatial resolution and the sensitivity to orbital patterns can be improved by functionalizing the STM tip (*6,13*). Furthermore, time resolution in STM-based orbital imaging has been achieved (*14*).

There is, however, a fundamental drawback of STM-based orbital imaging. The method is restricted to orbitals close to the chemical potential (Fermi energy $E_F$), typically in an energy range $E_F \pm 2$ eV. In many cases, however, it would be advantageous to be able to identify, analyze and image also deeper-lying orbitals. A case in point are σ-orbitals. Since their electron density is concentrated along the atom-atom interconnections in molecules, they play an important role in the local bonding between neighboring atoms. The capability to measure σ-orbitals as fingerprints of local chemical structures is therefore critical.

Photoemission orbital tomography (POT) (*7*) is a recent technique for the orbital analysis and orbital imaging of molecules on surfaces. For example, valence spectra have been deconvolved model-free into orbital-projected densities of states (*15,16*), orbitals have been reconstructed from experimental data in 2D and 3D (*17–19*), and orbital patterns in reciprocal space have been recorded with femtosecond time resolution (*20*). In POT, photoelectrons are collected in the half-space above the sample. The measured angular distribution of photoelectrons is governed by the spatial distribution of electrons in the initial state – the orbital. There is overwhelming evidence that for π-orbitals in conjugated molecules this relation is a straightforward Fourier transform (*21,22*). This is valid if the final state of the photoelectron after photoemission can be represented by a plane wave.

Since it is based on angle-resolved photoemission spectroscopy (ARPES), POT is not limited to a certain binding energy range if the photon energy is large enough. However, there is an important caveat. So far, the plane wave approximation has only been confirmed experimentally for π-orbitals close to $E_F$. Conceptually, the argument is based on the photoemission from identical orbitals (e.g., $p_z$-orbitals) on a set of identical atoms – and on the absence of appreciable photoelectron scattering, a good approximation for light atoms such as carbon, nitrogen and oxygen and relatively low photoelectron kinetic energies $E_{kin}$ (*22*). Then, the plane-wave approximation (PWA) becomes equivalent to the independent atomic center approximation (*23*). However, essential aspects of this argument may not hold for σ-orbitals, which derive from *s*-, $p_x$- and $p_y$-orbitals and have both larger binding energies and larger wave-vectors.

Here, we address the question whether the angular distributions of σ-orbitals can be explained in terms of the PWA. If this was the case, POT could straightforwardly be extended to the identification, analysis and imaging of σ-orbitals and their chemical phenomenology. This would open a much more direct way to study chemical bonding than the commonly used x-ray photoelectron spectroscopy, which provides indirect information on local bonding by measuring the influence of these bonds on the binding energy and spectral shapes of atomic core levels. With σ-based POT, one would gain the possibility to access the bonding orbitals directly.

Our model is the on-surface dehalogenation and cyclodehydrogenation of 10,10′-dibromo-9,9′-bianthracene (DBBA). DBBA and similar molecules are often employed to create carbon-based nanostructures by on-surface synthesis (*24–26*). The initially dehalogenated positions of the precursor can directly engage in an Ullmann-type C–C coupling (*27,28*), but can also become metalated or hydrogenated. The analytical discrimination between these competing reactions and products is important and we used it as a test case for POT on σ-orbitals.

We deposited DBBA (Fig. 1, **1**) on Cu(110) (*12,29*). DBBA undergoes catalytic dehalogenation and cyclodehydrogenation reactions at 525 K, during which the adsorbate planarizes and the π-conjugation expands to the entire fused carbon backbone, leading to the formation of a bisanthene-like species – possible products are shown in Fig. 1. While the possibility of a direct chemical bonding between the underlying substrate atoms and the carbon atoms in the 10,10′ positions (Fig. 1, **2**) (*29*), or even all carbon atoms along the zig-zag edge of bisanthene (Fig. 1, **3**), was excluded in a previous study based on the analysis of frontier π-orbitals (*12*), the discrimination between bisanthene (Fig. 1, **4**) and metalated bisanthene (Fig. 1, **5**) has not been achieved yet. Several examples of metal-molecule complexes were reported in similar reactions and hence it is not clear how the valency of carbon atoms in question will be saturated. Both hydrogen and copper adatoms are abundantly available and mobile at elevated temperatures in the reaction environment on the surface.

In a first attempt to discern between **4** and **5** we computed the adsorption structures and electronic properties of both species on Cu(110) by van-der-Waals-corrected density functional theory (DFT). We found that neither the projected density of states (pDOS) in the energy window between the Fermi energy and the onset of the Cu *d*-band nor the simulated photoemission $\mathbf{k}_\parallel$-maps for the lowest unoccupied (LUMO) and highest occupied (HOMO) molecular orbitals could discriminate between the two scenarios when compared to the experiment (Fig. S1). Note that upon adsorption on Cu(110), the bisanthene LUMO becomes occupied by charge transfer from the metal, making it accessible to POT. However, Fig. 2**A** shows that one of the two uppermost σ-orbitals is strongly affected by the replacement of the hydrogen in the 10,10′ positions by Cu atoms, leading to the expectation that **4** and **5** could possibly be discriminated by measuring σ-orbitals.

To check whether σ-orbitals are detectable in POT, we recorded photoemission band maps in two azimuthal directions up to a binding energy of $E_b \simeq 13$ eV (Fig. 2**B**,**C**). We know from previous work (*12,29*) that the product of the DBBA dehalogenation and cyclohydrogenation is orientated with its zig-zag edges along the substrate [001] direction and with the armchair edges along [1$\bar{1}$0]. In addition to the frontier π-orbitals, one can clearly identify in Fig. 2**B**,**C** several molecular emissions below the *d*-band: a π-band dispersing between $E_b \simeq 4$ to 10 eV at $|\mathbf{k}_\parallel| \leq 1.5$ Å$^{-1}$, and a band between $E_b \simeq 5$ to 13 eV at higher $|\mathbf{k}_\parallel|$. The large $|\mathbf{k}_\parallel|$ of the latter suggest that it might be of σ-character.

Figure 3 displays all calculated orbitals of bisanthene in the energy range from 3 to 13.5 eV. We arranged the $\mathbf{k}_\parallel$-maps, calculated by Fourier transforming the orbitals of the isolated molecule, according to their smallest-$|\mathbf{k}_\parallel|$ emission lobes; the latter roughly scale with the number of nodal planes along $k_x$ (zig-zag direction of bisanthene and [001] direction of Cu(110)) and $k_y$ (armchair direction and [1$\bar{1}$0]). We therefore labeled the orbitals according to their type (π or σ) and the number of nodal planes (0,1,2...) along $k_x$ and $k_y$. For instance, the HOMO π(2,3) exhibits two (three) nodal planes along $k_x$ ($k_y$), while the σ(7,3) orbital has seven and three nodes, respectively. π(4,0) is the LUMO.

Next, we compared the theoretical patterns in Fig. 3 to experimental $\mathbf{k}_\parallel$-maps (movie S1). This procedure is well-founded for π-orbitals, but still to be substantiated for σ-orbitals because of the uncertain applicability of the PWA for them. Of the 15 π- and 27 σ-orbitals in Fig. 3 we found 12 π- and 18 σ-orbitals in the range from $E_b = 0$ to 10eV. Only 9 σ-orbitals were not identified with certainty. We used a *k*-space deconvolution with theoretical $\mathbf{k}_\parallel$-maps for overlapping orbitals (Fig. S3–S7). The large number of identified bisanthene σ-orbitals in excellent agreement with theory confirms that POT can not only be applied to $p_z$-derived but also to *s*-, $p_x$- and $p_y$-derived orbitals. This is a remarkable finding for which no theoretical explanation has been given yet. Nevertheless, the experiment speaks for itself.

In the remainder of the paper, we focus on the two uppermost σ-orbitals just below the Cu *d*-band, using them to discriminate between products **4** and **5** of the surface reaction in Fig. 1. In Fig. 4**A** the experimental $\mathbf{k}_\parallel$-map at $E_b$=5.16eV is displayed. Its pattern is complex and shows emissions from several states: σ-orbitals and π-orbitals (circles), and high-$E_b$ tails of the Cu *d*-band (triangles). The strong emission lobes at $|\mathbf{k}_\parallel| \geq 2$ Å$^{-1}$ can be accounted for by the two uppermost σ-orbitals σ(7,3) (red) and σ(0,8) (blue), whose real-space distributions are displayed in Fig. 2**A**. As can be seen in their simulated *k*-space distributions (Fig. 4**B** inset), the emissions at $\mathbf{k}_\parallel = (0, \pm 2.89)$ Å$^{-1}$ arise from σ(0,8), while those at $(\pm 2.56, \pm 1.39)$ Å$^{-1}$ can be from both σ(0,8) and σ(7,3).

To quantitatively decompose the experimental $\mathbf{k}_\parallel$-maps, we minimized the sum of least squares (*15*)

$$\chi^2 = \int d\mathbf{k}_\parallel [I(E_{kin}, \mathbf{k}_\parallel) - \sum_i a_i(E_b) I_i(\mathbf{k}_\parallel)]^2 \quad (1)$$

to extract the $E_b$-dependent contributions $a_i$ of the component orbitals to the total POT intensity distribution $I(E_{kin}, \mathbf{k}_\parallel)$. $I_i(\mathbf{k}_\parallel)$ are the calculated $\mathbf{k}_\parallel$-maps of the orbitals of free bisanthene (Fig. 3). Fig. 4**B** shows the $a_i$ for three orbitals σ(0,8), σ(7,3) and π(0,3). They have the meaning of an experimentally-derived pDOS and can be compared to their theoretically calculated counterparts, to which we now turn.

In Fig. 4**C**,**E** we show simulated $\mathbf{k}_\parallel$-maps for bisanthene (**4**) and metalated bisanthene (**5**) on Cu(110). These maps have been calculated using a damped plane wave as the final state, which prevents the overrepresentation of bulk states from the substrate (*30*). For bisanthene, the six bright emission lobes at $|\mathbf{k}_\parallel| \geq 2$ Å$^{-1}$ (Fig. 4**C**) correctly reproduce the experimental pattern. In contrast, for metalated bisanthene, the four emission lobes of the σ(7,3) orbital dominate, while the emission from σ(0,8) is barely visible (Fig. 4**E**). This already leads us to conclude that bisanthene agrees better with the experimental data and the metalation with Cu adatoms is unlikely. This conclusion is further corroborated by comparing the calculated pDOS in Fig. 4**D**,**F** to the experimental one in Fig. 4**B**. The former are obtained by calculating the densities of states for the complete molecule/substrate systems and projecting them on the σ(0,8) and σ(7,3) orbitals.

For bisanthene, the pDOS for both orbitals exhibit well-defined peaks with similar intensities and energies, therein resembling closely the experiment. In contrast, the σ(0,8) pDOS for metalated bisanthene is spread out broadly in energy with its maxima well-separated from that of σ(7,3). This indicates that the σ(0,8) orbital of metalated bisanthene hybridizes strongly with the substrate, presumably through the Cu atoms, which are fully integrated into the lobe structure of σ(0,8) (but, notably, not of σ(7,3)) and which upon adsorption couple strongly to the metal surface. Because this hybridization is not observed in the experimental pDOS (Fig. 4**B**), we can definitively rule out the metalated product **5**. To cross check, we repeated the deconvolution of the experimental $I(E_{kin}, \mathbf{k}_\parallel)$ with theoretical $\mathbf{k}_\parallel$-maps $I_i(\mathbf{k}_\parallel)$ of free metalated bisanthene (Fig. S9). This delivers a very similar result to the experimental pDOS in Fig. 4**B**, which, however, is inconsistent with the calculated pDOS for the metalated species. Thus, only bisanthene as a product of the surface reaction allows a consistent interpretation of our σ-orbital data. The analysis of the deep-lying π(0,3) orbital is in full accord with this result (Supplement).

In conclusion, we advanced photoemission orbital tomography, first by substantially extending the accessible binding energy range, in particular far below the *d*-band of the substrate, and second by successfully detecting σ-orbitals. We found that the PWA for the final state is also suited for σ-orbitals. For the product of the catalytic dehalogenation and cyclodehydrogenation of DBBA on Cu(110), we demonstrated that almost the complete orbital spectrum in a wide binding energy range is detectable. Since – depending on orbital shape, adsorption site, and registry with the sample – it is usually only a small number of orbitals which provide key information regarding chemical modifications, the ability to image the complete spectrum of valence orbitals with photoemission orbital tomography in an unrivaled binding energy range provides an invaluable advantage for the analysis of surface chemical reactions.

**Acknowledgments**

We thank Philipp Hurdax (Karl-Franzens-Universität Graz, Austria), Hendrik Kaser (Physikalisch-Technische Bundesanstalt, Germany) and John Riley (La Trobe University, Australia) for experimental support.

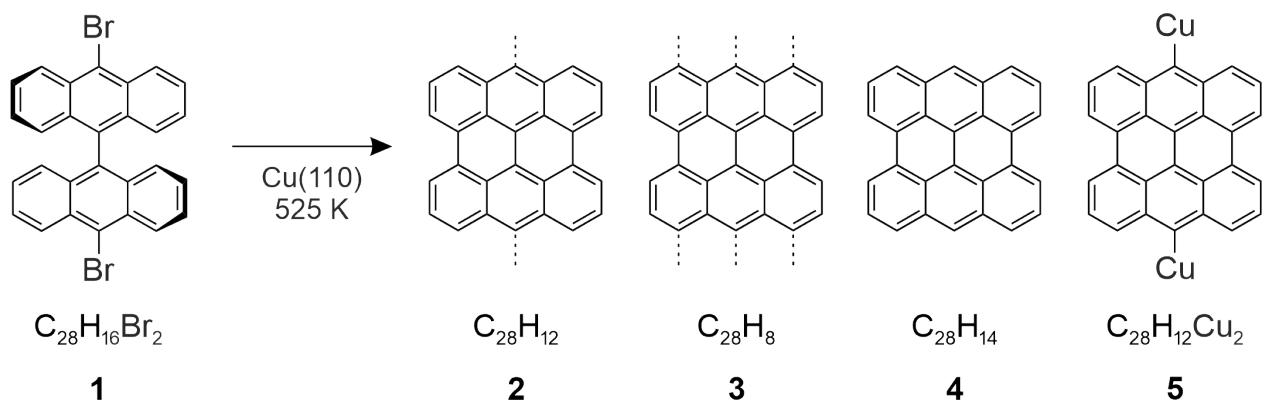

**Fig. 1. DBBA and possible reaction products on Cu(110).** Dotted lines indicate bonds between carbon atoms and Cu atoms in the substrate.

**Fig. 2. σ-orbitals and ARPES band maps.** (**A**) σ(7,3) and σ(0,8) orbitals of bisanthene ($C_{28}H_{14}$, top) and $C_{28}H_{12}Cu_2$ (bottom). (**B,C**) Band maps along the [$1\bar{1}0$] and [001] directions. π- and σ-bands are labeled. The white dashed lines denote $E_b$ of the $\mathbf{k}_\parallel$-map in Fig. 4A.

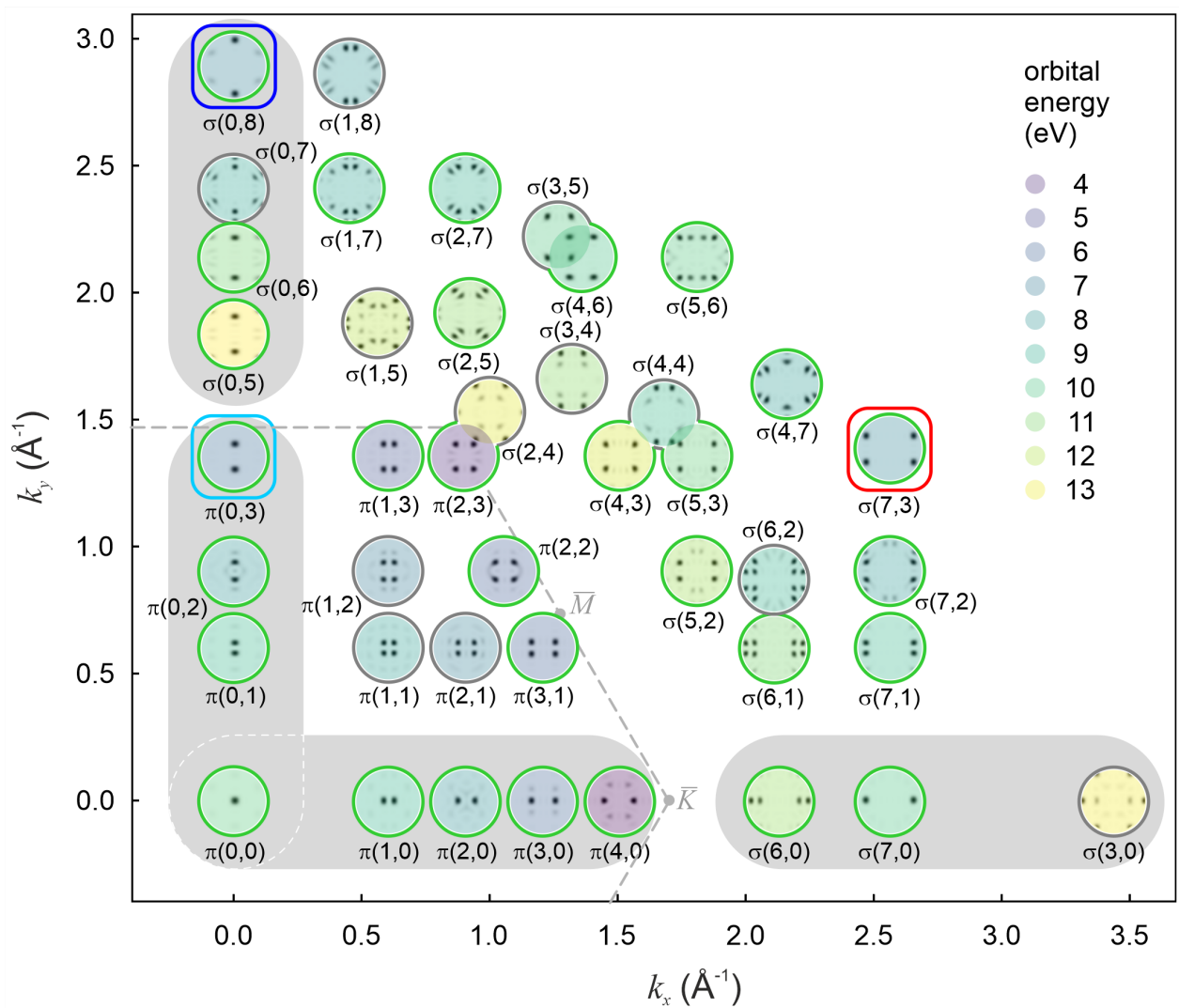

**Fig. 3. Orbitals of bisanthene.** The DFT-calculated $\mathbf{k}_\parallel$-maps of the free-molecule orbitals arranged by ($k_x$, $k_y$) of their smallest-$|\mathbf{k}_\parallel|$ emission lobes. Orbital labels indicate the number of nodal planes along the two principal directions. Calculated orbital energies (zero at vacuum level) are indicated by color. Circles indicate whether these orbitals have been identified (green) or not identified (grey) in the experimental dataset. The grey-shaded areas denote orbitals of the π and σ-bands marked in Fig. 2**B**,**C**. The dashed line marks the Brillouin zone of graphene. Cf. Fig. S2 for a plot with the images of the orbitals.

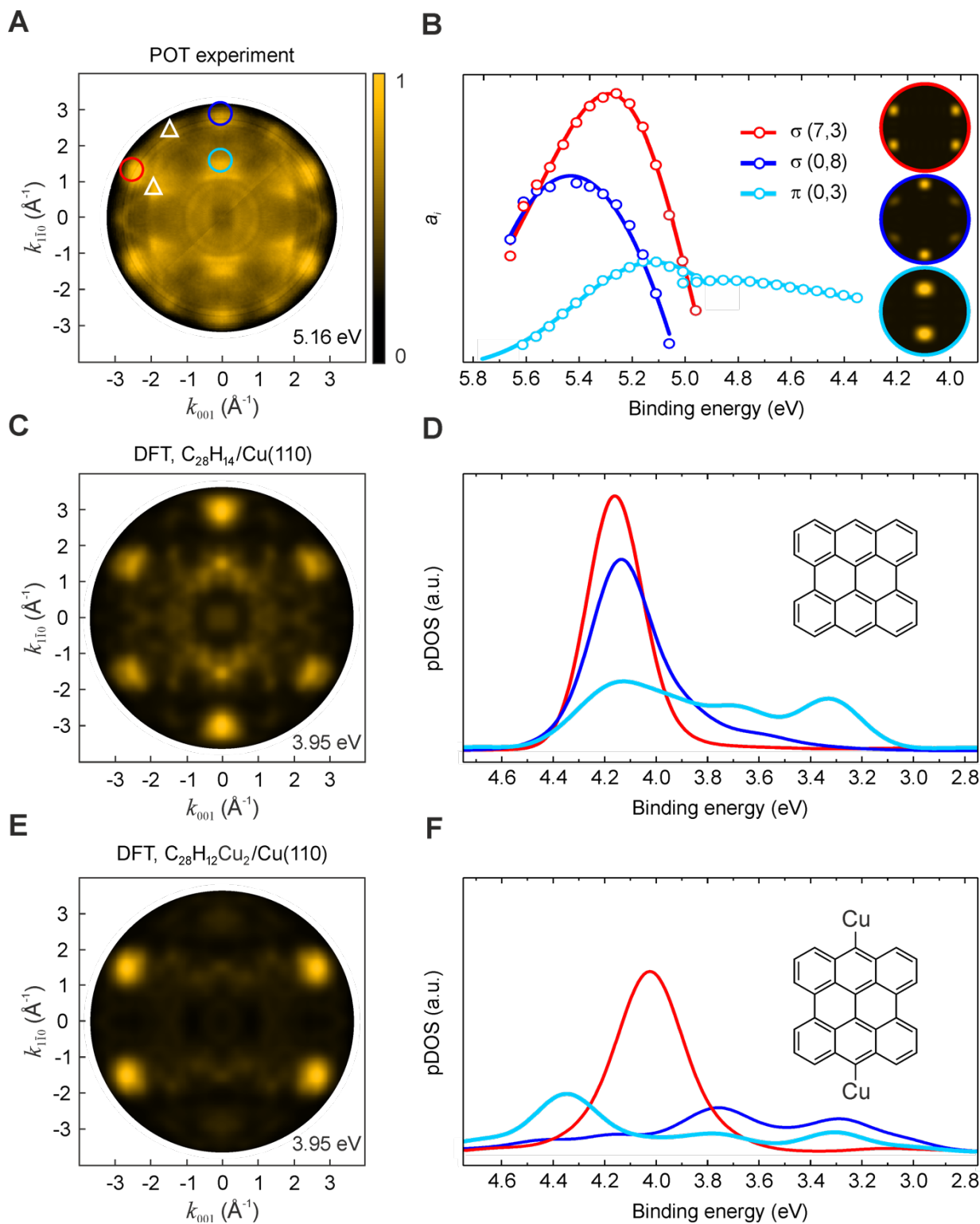

**Fig. 4. $k_\parallel$-maps and pDOS for two possible reaction products.** (**A**) $k_\parallel$-map measured at $E_b$ = 5.16eV. The circles denote emissions from molecular orbitals (color as in **B**,**D** and **F**), the triangles from metal states. The map of clean Cu(110) is shown in Fig. S8.
(**B**) Experimental pDOS of σ(7,3), σ(0,8) and π(0,3) orbitals. The data points were obtained from the deconvolution of an $I(E_{kin}, k_\parallel)$ data cube, using the theoretical $k_\parallel$-maps of free bisanthene (cf. inset). For π(0,3), two adjacent datasets with a $E_b$ range of 1 eV each are displayed. (**C**) Theoretical $k_\parallel$-map of $C_{28}H_{14}$/Cu(110) at the calculated $E_b$ = 3.95 eV.
(**D**) Theoretical pDOS of σ(7,3), σ(0,8) and π(0,3) orbitals of $C_{28}H_{14}$/Cu(110). (**E**) Theoretical

$k_\parallel$-map of $C_{28}H_{12}Cu_2$/Cu(110) at the calculated $E_b$ = 3.95 eV. (**F**) Theoretical pDOS of $\sigma(7,3)$, $\sigma(0,8)$ and $\pi(0,3)$ orbitals of $C_{28}H_{12}Cu_2$/Cu(110).

Supplementary Materials for

# Momentum-space imaging of σ-orbitals for chemical analysis

Anja Haags[1,2,3], Xiaosheng Yang[1,2,3], Larissa Egger[4], Dominik Brandstetter[4], Hans Kirschner[5], François C. Bocquet[1,2], Georg Koller[4], Alexander Gottwald[5], Mathias Richter[5], J. Michael Gottfried[6], Michael G. Ramsey[4], Peter Puschnig[4]*, Serguei Soubatch[1,2]*, F. Stefan Tautz[1,2,3]

[1]Peter Grünberg Institut (PGI-3), Forschungszentrum Jülich; Jülich, Germany.

[2]Jülich Aachen Research Alliance (JARA), Fundamentals of Future Information Technology; Jülich, Germany.

[3]Experimentalphysik IV A, RWTH Aachen University; Aachen, Germany.

[4]Institut für Physik, Karl-Franzens-Universität Graz, NAWI Graz; Graz, Austria.

[5]Physikalisch-Technische Bundesanstalt (PTB); Berlin, Germany.

[6]Fachbereich Chemie, Philipps-Universität Marburg; Marburg, Germany.

*Corresponding author. Email: peter.puschnig@uni-graz.at, s.subach@fz-juelich.de

**Sample preparation**

The preparation of bisanthene/Cu(110) and the photoemission experiments were performed in ultra-high vacuum ($p \simeq 10^{-10}$ mbar). The Cu(110) single crystal was cleaned by several cycles of sputtering by $Ar^+$-ions at 1 keV and subsequent annealing at 800 K. To prepare a monolayer of bisanthene, a film of the 10,10′-dibromo-9,9′-bianthracene (DBBA) precursor (Sigma-Aldrich, CAS number 121848-75-7) was deposited by evaporation from a molecular evaporator (Kentax GmbH) onto the crystal surface held at room temperature. Subsequently, the sample was annealed at 525 K to trigger the surface chemical reaction.

**Photoemission experiments**

Photoemission experiments were conducted at the Metrology Light Source insertion device beamline of the Physikalisch-Technische Bundesanstalt (Berlin, Germany) (*31*). p-polarized ultraviolet light ($h\nu$=57 eV) with an incidence angle of 40° to the surface normal was used. In this geometry, the **A**∥**k** condition (*22*) is approximately fulfilled for most molecular emissions in forward direction. Two different types of photoemission experiments were conducted using the toroidal electron analyzer (*32*). To obtain the band maps, the photoemission intensity was recorded in the emission angle range from -85° to +85° along the [1$\bar{1}$0] and [001] directions of Cu(110) as a function of kinetic energy $E_{kin}$. Experimental $\mathbf{k}_\parallel$-maps at a fixed binding energy $E_b$ were obtained from the three-dimensional data cube $I(E_{kin}, \mathbf{k}_\parallel)$ – intensity of photoemission as a function of kinetic energy $E_{kin}$ and parallel momentum $\mathbf{k}_\parallel$ – recorded while rotating the sample around its normal in 1° steps. In this way, the full photoemission intensity distribution in the $\mathbf{k}_\parallel$ plane perpendicular to the sample normal was recorded.

**Density functional theory calculations**

The geometry and electronic structure of the free and surface-adsorbed molecules were calculated in the framework of density functional theory (DFT) with the generalized gradient approximation according to Perdew, Burke and Ernzerhof (PBE-GGA) (*33*) for the exchange-correlation potential. The program packages NWChem (*34*) (for free molecules) and VASP (*35,36*) (for surface-adsorbed molecules) were used. For $C_{28}H_{14}$/Cu(110) and $C_{28}H_{12}Cu_2$/Cu(110), the D3 correction (*37*) was applied to account for the van-der-Waals interaction. The projector augmented wave (PAW) method (*38*) was employed with a plane wave cutoff of 500 eV. Structures were optimized on a Monkhorst-Pack 3×3×1 grid of *k*-points with a first-order Methfessel-Paxton smearing of 0.2 eV. Using a repeated slab approach, the Cu(110) substrate was modeled with 5 atomic layers, a lattice parameter of $a = 3.61$ Å and a vacuum layer of Cu at least 17 Å between the slabs to avoid spurious electric fields (*39*). The correct adsorption site for each molecule has been determined by testing several high-symmetry adsorption sites (hollow, top, short bridge and long bridge) in a local geometry optimization approach, allowing all molecular degrees of freedom and the topmost two Cu-layers to relax until forces were below 0.01 eV/Å. The orbital $\mathbf{k}_\parallel$-maps were computed within the one-step model of photoemission and the plane wave final state approximation (*7*) using an increased *k*-grid of 5×5×3. To account for the mean free path of the photoemitted electrons, an exponential damping factor in the final state of 0.5 Å$^{-1}$ between substrate and molecule was introduced (*40*).

**pDOS of the π(0,3) Orbital**

The experimental $\mathbf{k}_\parallel$-maps and pDOS in Fig. 4**A**,**B** also exhibit contributions of the π(0, 3) orbital (light blue circle/light blue curve). Unlike the two σ-orbitals σ(7,3) and σ(0,8), π(0,3) is strongly spread out in energy. In fact, the calculated π(0,3) pDOS of bisanthene on Cu(110) resembles the experimental result much better than metalated bisanthene does, providing yet another piece of evidence in favor of bisanthene as a product of the surface reaction. Incidentally, the strong hybridization of π(0,3) reveals an intriguing insight into the chemistry at the interface between bisanthene and Cu(110): the lobes of π(0,3), located on the zig-zag chains of bisanthene, perfectly match the periodicity of the substrate lattice in the [1$\bar{1}$0] direction and bridge three Cu atomic rows in the [001] direction (Fig. S10). This assists the hybridization between the π(0,3) orbital and the Cu(110) surface, leading to its strong broadening in Fig. 4**B**. Because metalated bisanthene occupies another adsorption site (Fig. S11), its π(0,3) orbital cannot hybridize with the substrate in the same way, resulting in a different pDOS with the leading feature at larger binding energies than σ(0,8) and σ(7,3) (Fig. 4**F**). The analysis of the deep-lying π(0,3) orbital is thus in full accord with the results from the uppermost σ-orbitals and confirms the viability of σ-orbital POT for chemical analysis.

**Supplementary References**

**Supplementary Figures**

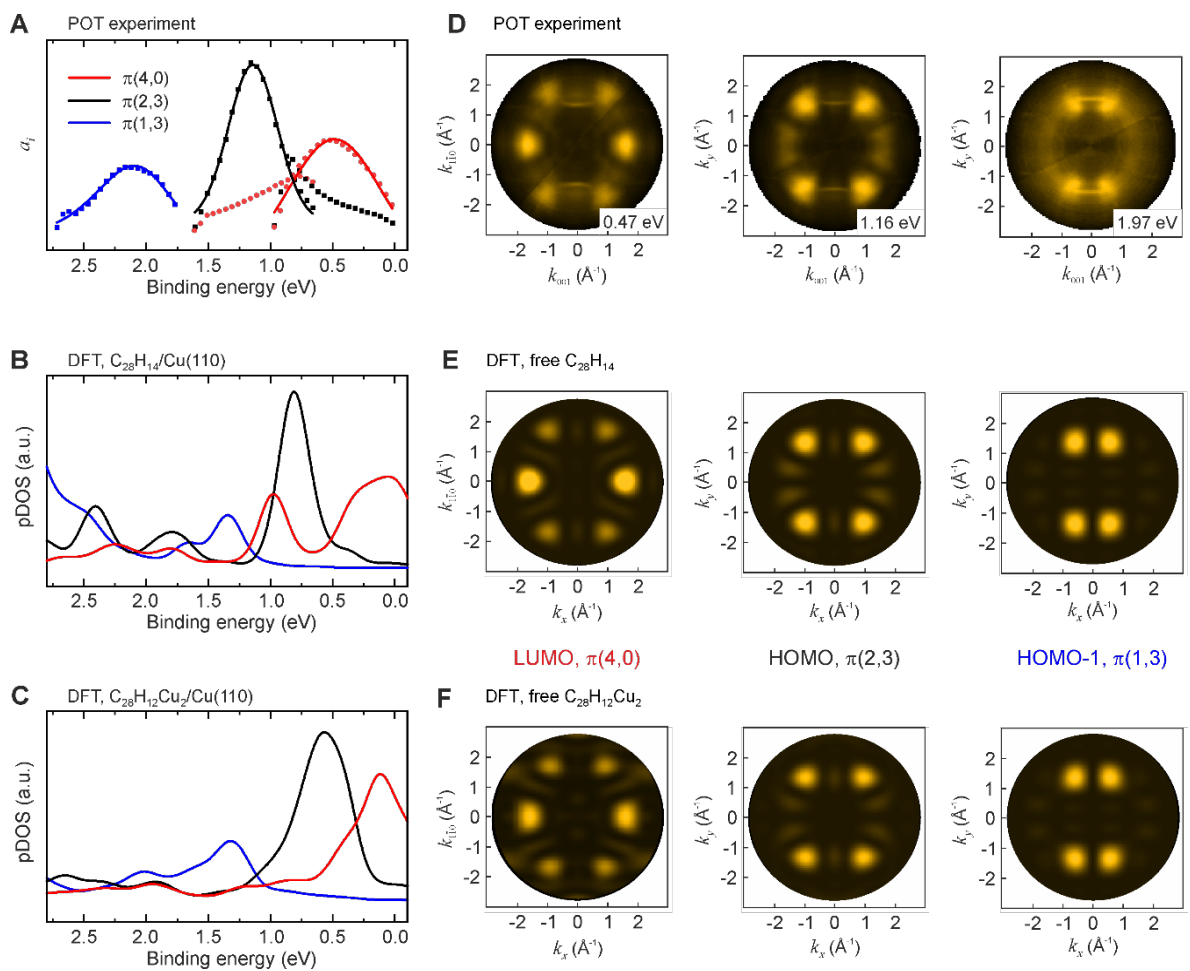

**Fig. S1. Frontier π-orbitals of bisanthene.** (**A**) Experimental projected density of states (pDOS) obtained by deconvolution of the photoemission orbital tomography (POT) results using the $k_\parallel$-maps of free $C_{28}H_{14}$ shown in **E**. Note that deconvolution using the $k_\parallel$-maps of free $C_{28}H_{12}Cu_2$ shown in **F** delivers qualitatively similar results. (**B**) Theoretical pDOS of $C_{28}H_{14}$/Cu(110) calculated by DFT. (**C**) Theoretical pDOS of $C_{28}H_{12}Cu_2$/Cu(110) calculated by DFT. (**D**) Experimental $k_\parallel$-maps recorded at different binding energies $E_b$. (**E**) Theoretical $k_\parallel$-maps calculated for LUMO, HOMO and HOMO-1 of free $C_{28}H_{14}$. (**F**) Theoretical $k_\parallel$-maps calculated for LUMO, HOMO and HOMO-1 of free $C_{28}H_{12}Cu_2$.

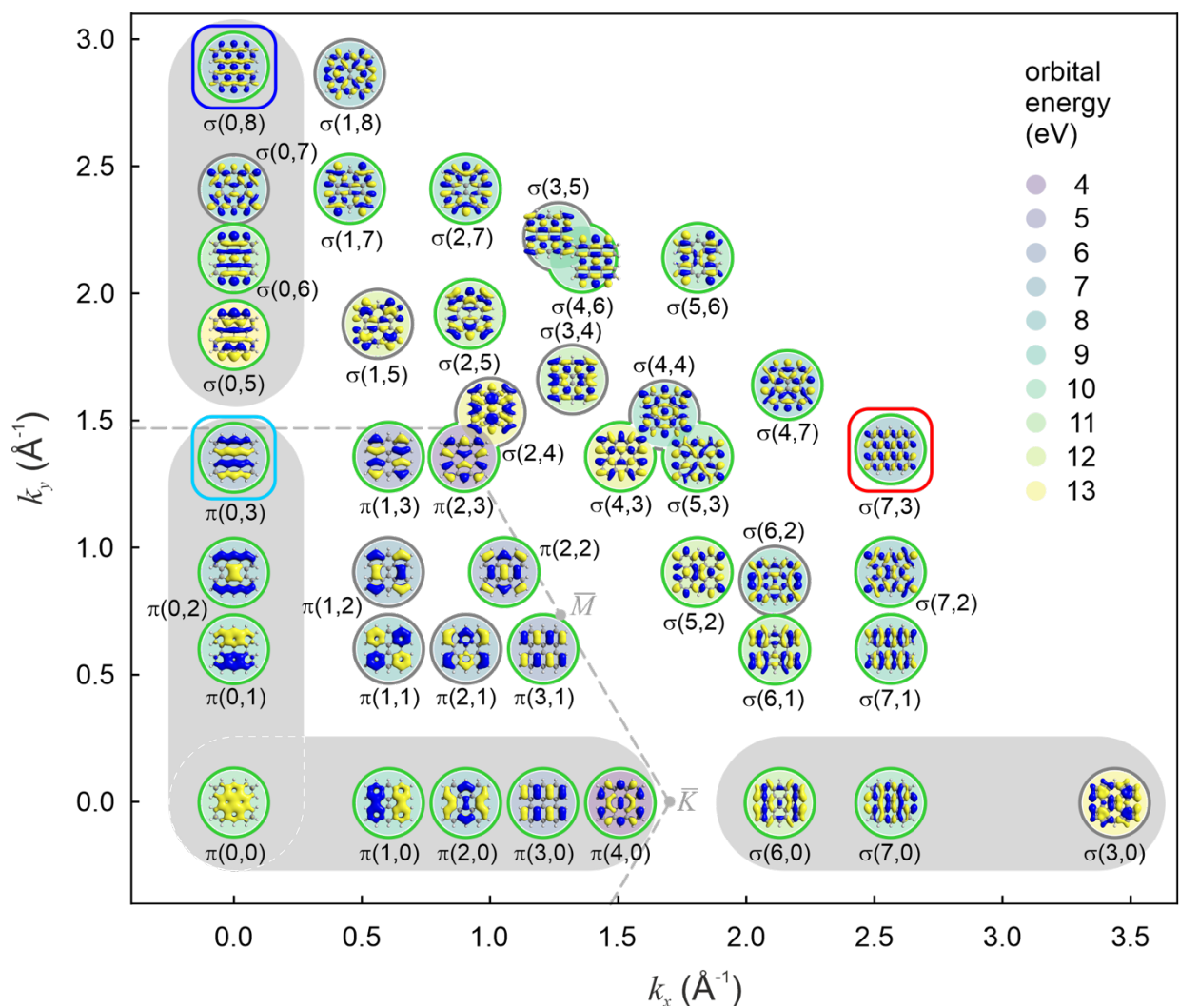

**Fig. S2. Orbitals of bisanthene.** The DFT-calculated orbitals of the free molecule are arranged according to the ($k_x$, $k_y$)-positions of their smallest $|\mathbf{k}_\parallel|$ emission lobes in their respective $\mathbf{k}_\parallel$- maps. The orbitals are labeled according to the number of nodal planes along the two principal directions (cf. main text). Calculated orbital energies with respect to the vacuum level are indicated by the color scale. Circles around the orbitals indicate whether they have been identified (green) or not identified (grey) in the experimental dataset. Red, blue and light blue frames mark the σ(7,3), σ(0,8) and π(0,3) orbitals, respectively. The grey shaded areas denote orbitals of the π and σ-bands marked in Fig. 2**B**,**C**. As a reference length scale in reciprocal space, the dashed grey line marks the Brillouin zone of graphene.

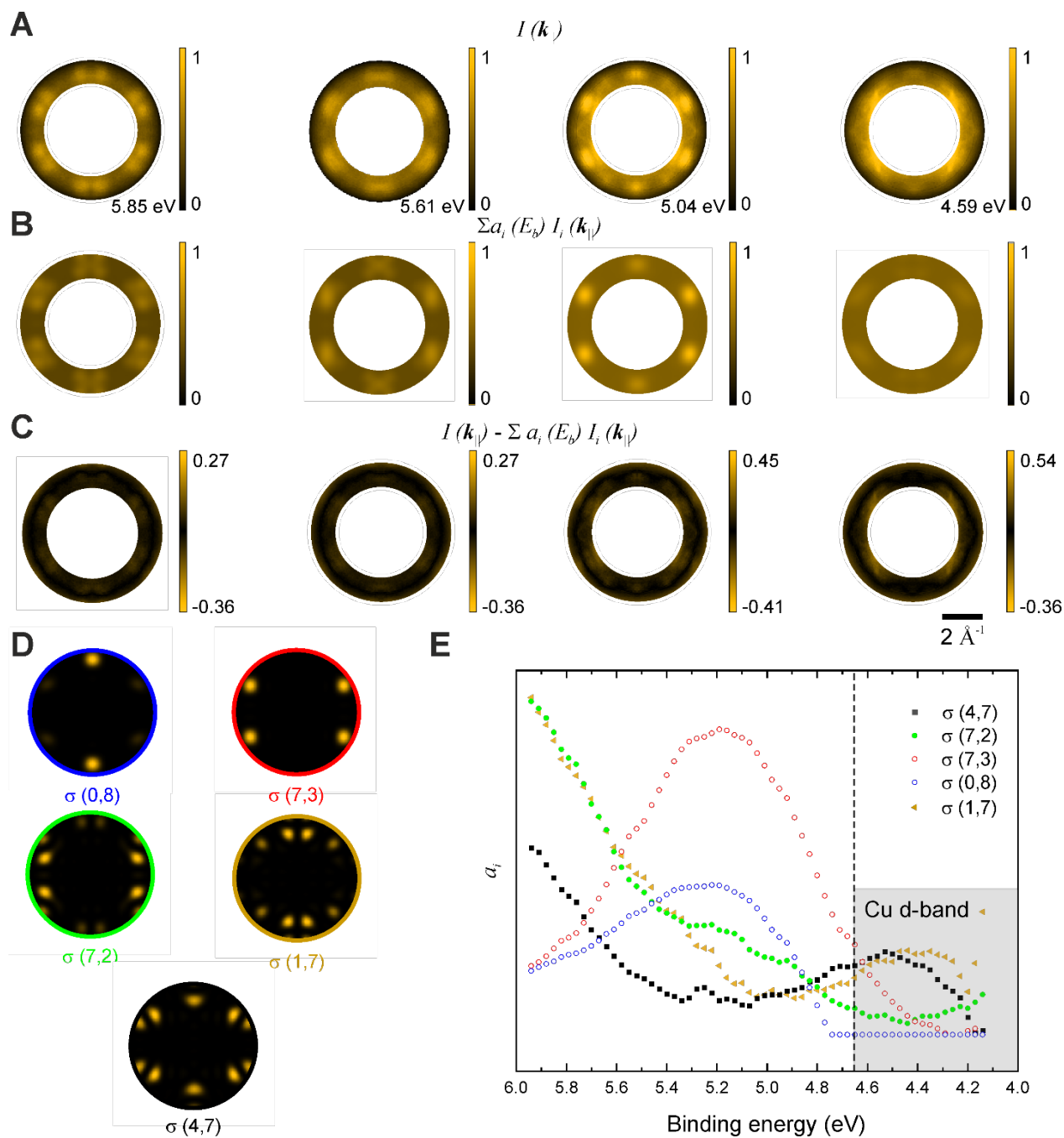

**Fig. S3. Deconvolution of experimental POT results in $E_b$ range of 4.15–6.00 eV.**
(**A**) Experimental $\mathbf{k}_\parallel$-maps $I(\mathbf{k}_\parallel)$ measured at corresponding binding energies. (**B**) $\mathbf{k}_\parallel$-maps $\sum_i a_i(E_b)I_i(\mathbf{k}_\parallel)$ constructed as intensity sum of theoretical $\mathbf{k}_\parallel$-maps $I_i(\mathbf{k}_\parallel)$ of free $C_{28}H_{14}$ shown in **D** and weighted using the fitting parameters $a_i$ from **E**. (**C**) $\mathbf{k}_\parallel$-maps of the deconvolution residual $I(\mathbf{k}_\parallel)-\sum_i a_i(E_b)I_i(\mathbf{k}_\parallel)$. (**D**) Theoretical $\mathbf{k}_\parallel$-maps $I_i(\mathbf{k}_\parallel)$ of free $C_{28}H_{14}$ of the σ-orbitals used for deconvolution. (**E**) Fitting parameters $a_i$ – the experimental pDOS of the σ-orbitals obtained from deconvolution. All $\mathbf{k}_\parallel$-maps are restricted to $|\mathbf{k}_\parallel| \geq 2.2$ Å$^{-1}$ to exclude low $|\mathbf{k}_\parallel|$ emissions from π states.

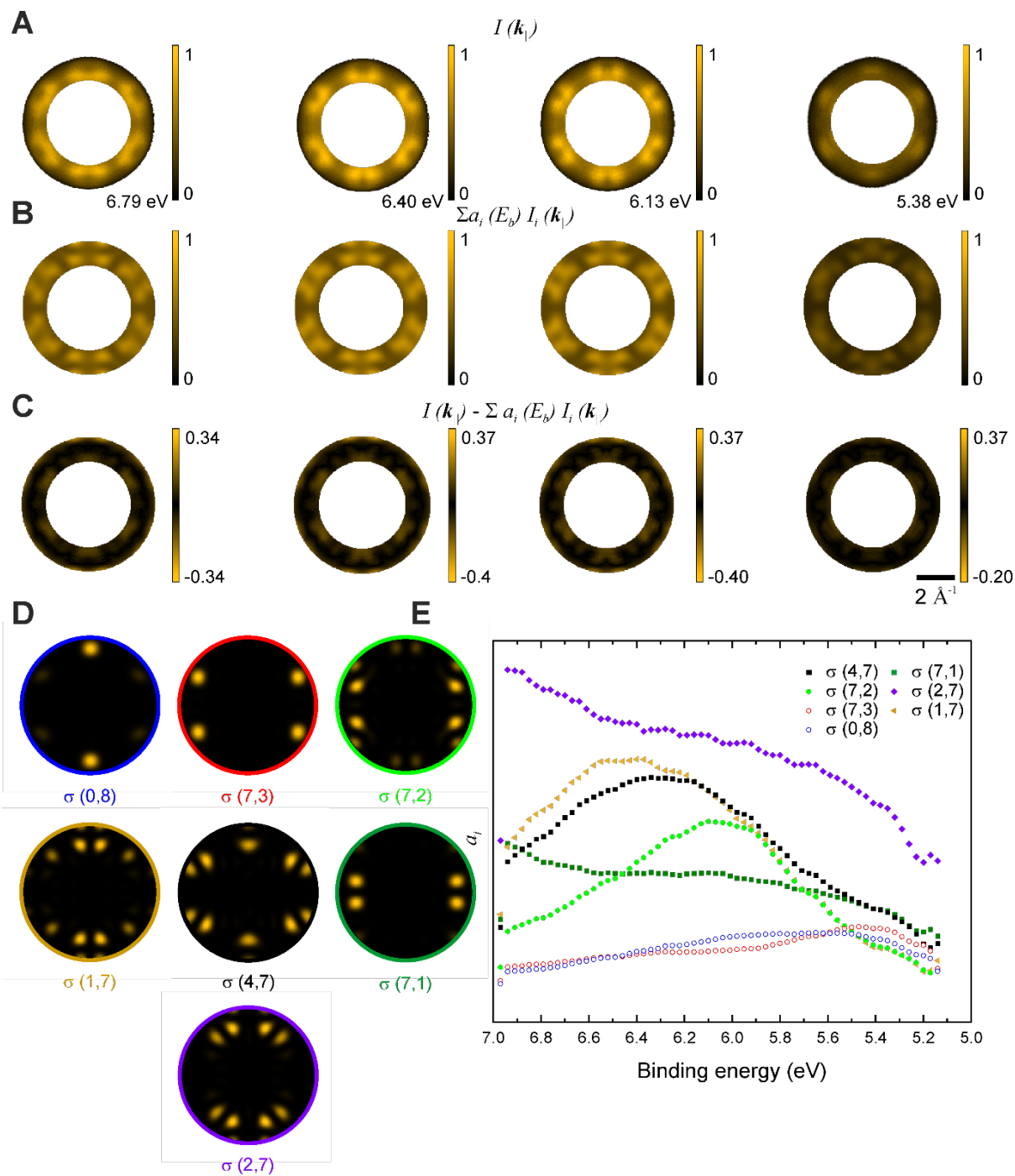

**Fig. S4. Deconvolution of experimental POT results in $E_b$ range of 5.15–7.00 eV.**
(**A**) Experimental $\mathbf{k}_\parallel$-maps $I(\mathbf{k}_\parallel)$ measured at corresponding binding energies. (**B**) $\mathbf{k}_\parallel$-maps $\sum_i a_i(E_b)I_i(\mathbf{k}_\parallel)$ constructed as intensity sum of theoretical $\mathbf{k}_\parallel$-maps $I_i(\mathbf{k}_\parallel)$ of free $C_{28}H_{14}$ shown in **D** and weighted using the fitting parameters $a_i$ from **E**. (**C**) $\mathbf{k}_\parallel$-maps of the deconvolution residual $I(\mathbf{k}_\parallel)-\sum_i a_i(E_b)I_i(\mathbf{k}_\parallel)$. (**D**) Theoretical $\mathbf{k}_\parallel$-maps $I_i(\mathbf{k}_\parallel)$ of free $C_{28}H_{14}$ of the σ-orbitals used for deconvolution. (**E**) Fitting parameters $a_i$ – the experimental pDOS of the σ-orbitals obtained from deconvolution. All $\mathbf{k}_\parallel$-maps are restricted to $|\mathbf{k}_\parallel| \geq 2.2$ Å$^{-1}$ to exclude low $|\mathbf{k}_\parallel|$ emissions from π states.

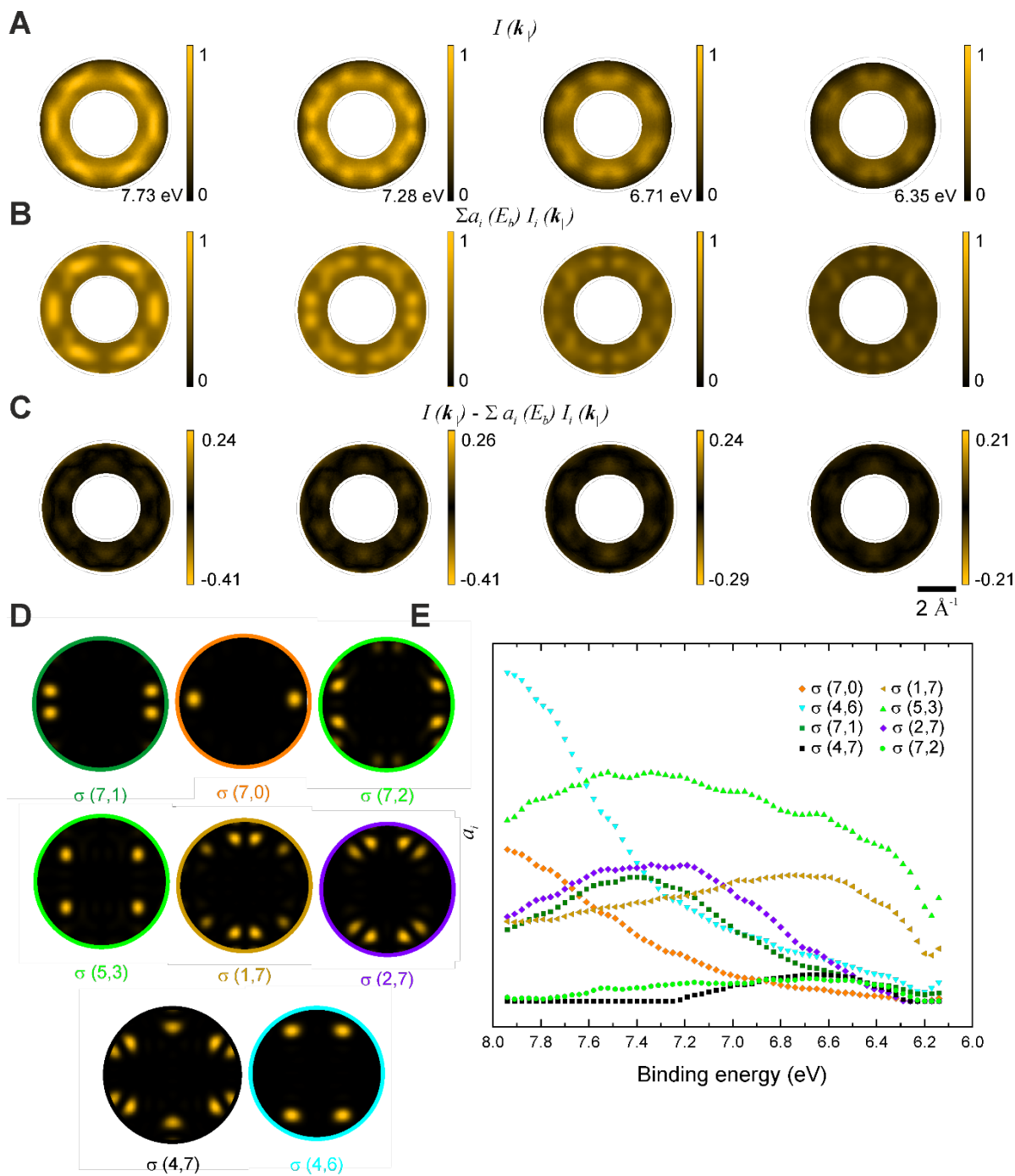

**Fig. S5. Deconvolution of experimental POT results in $E_b$ range of 6.15–8.00 eV.**
(**A**) Experimental $k_\parallel$-maps $I(k_\parallel)$ measured at corresponding binding energies. (**B**) $k_\parallel$-maps $\sum_i a_i(E_b) I_i(k_\parallel)$ constructed as intensity sum of theoretical $k_\parallel$-maps $I_i(k_\parallel)$ of free $C_{28}H_{14}$ shown in **D** and weighted using the fitting parameters $a_i$ from **E**. (**C**) $k_\parallel$-maps of the deconvolution residual $I(k_\parallel) - \sum_i a_i(E_b) I_i(k_\parallel)$. (**D**) Theoretical $k_\parallel$-maps $I_i(k_\parallel)$ of free $C_{28}H_{14}$ of the σ-orbitals used for deconvolution. (**E**) Fitting parameters $a_i$ – the experimental pDOS of the σ-orbitals obtained from deconvolution. All $k_\parallel$-maps are restricted to $|k_\parallel| \geq 1.6$ Å$^{-1}$ to exclude low $|k_\parallel|$ emissions from π states.

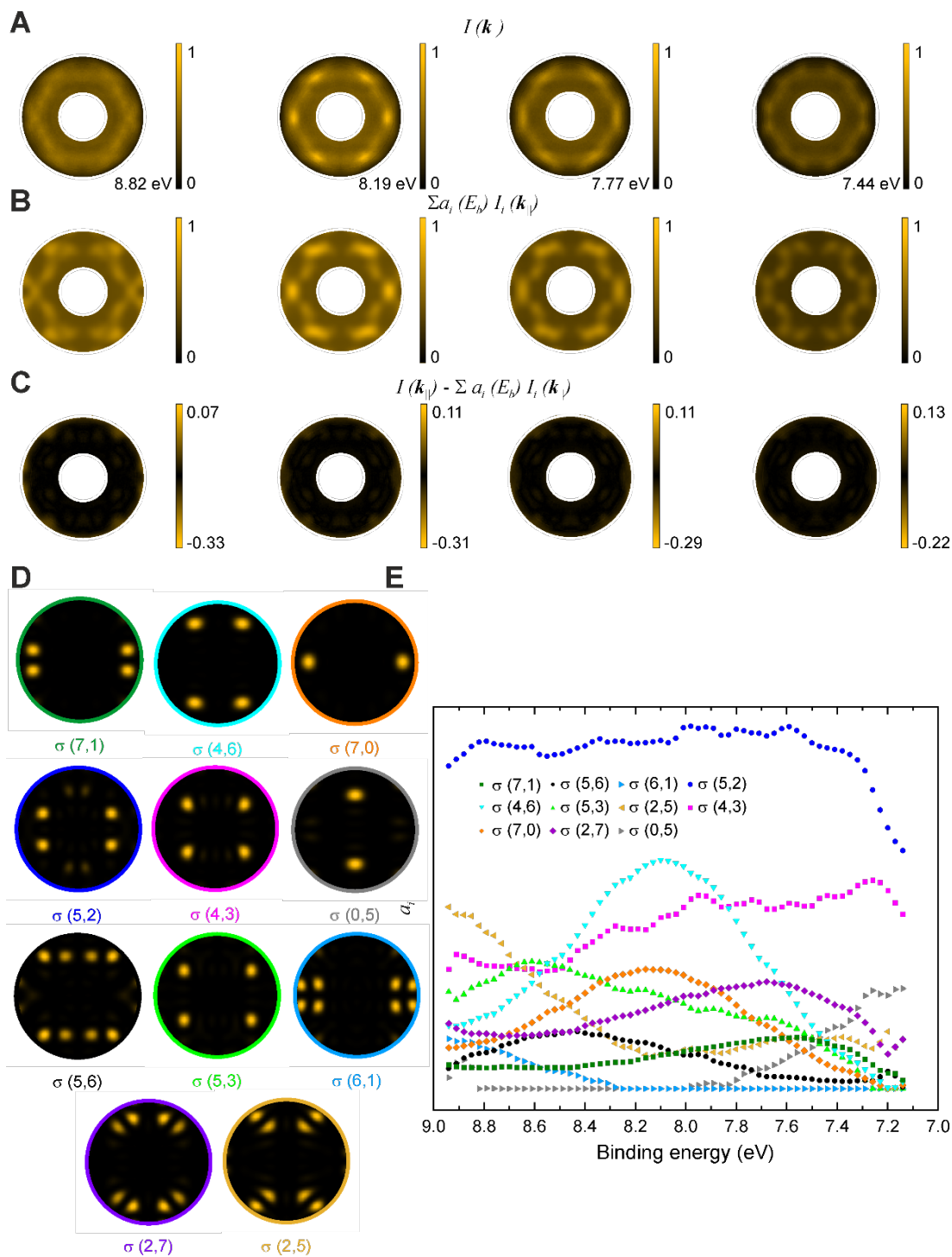

**Fig. S6. Deconvolution of experimental POT results in $E_b$ range of 7.15–9.00 eV.**
(**A**) Experimental $\mathbf{k}_\parallel$-maps $I(\mathbf{k}_\parallel)$ measured at corresponding binding energies. (**B**) $\mathbf{k}_\parallel$-maps $\sum_i a_i(E_b) I_i(\mathbf{k}_\parallel)$ constructed as intensity sum of theoretical $\mathbf{k}_\parallel$-maps $I_i(\mathbf{k}_\parallel)$ of free $C_{28}H_{14}$ shown in **D** and weighted using the fitting parameters $a_i$ from **E**. (**C**) $\mathbf{k}_\parallel$-maps of the deconvolution residual $I(\mathbf{k}_\parallel) - \sum_i a_i(E_b) I_i(\mathbf{k}_\parallel)$. (**D**) Theoretical $\mathbf{k}_\parallel$-maps $I_i(\mathbf{k}_\parallel)$ of free $C_{28}H_{14}$ of the σ-orbitals used for deconvolution. (**E**) Fitting parameters $a_i$ – the experimental pDOS of the σ-orbitals obtained from deconvolution. All $\mathbf{k}_\parallel$-maps are restricted to $|\mathbf{k}_\parallel| \geq 1.3$ Å$^{-1}$ to exclude low $|\mathbf{k}_\parallel|$ emissions from π states.

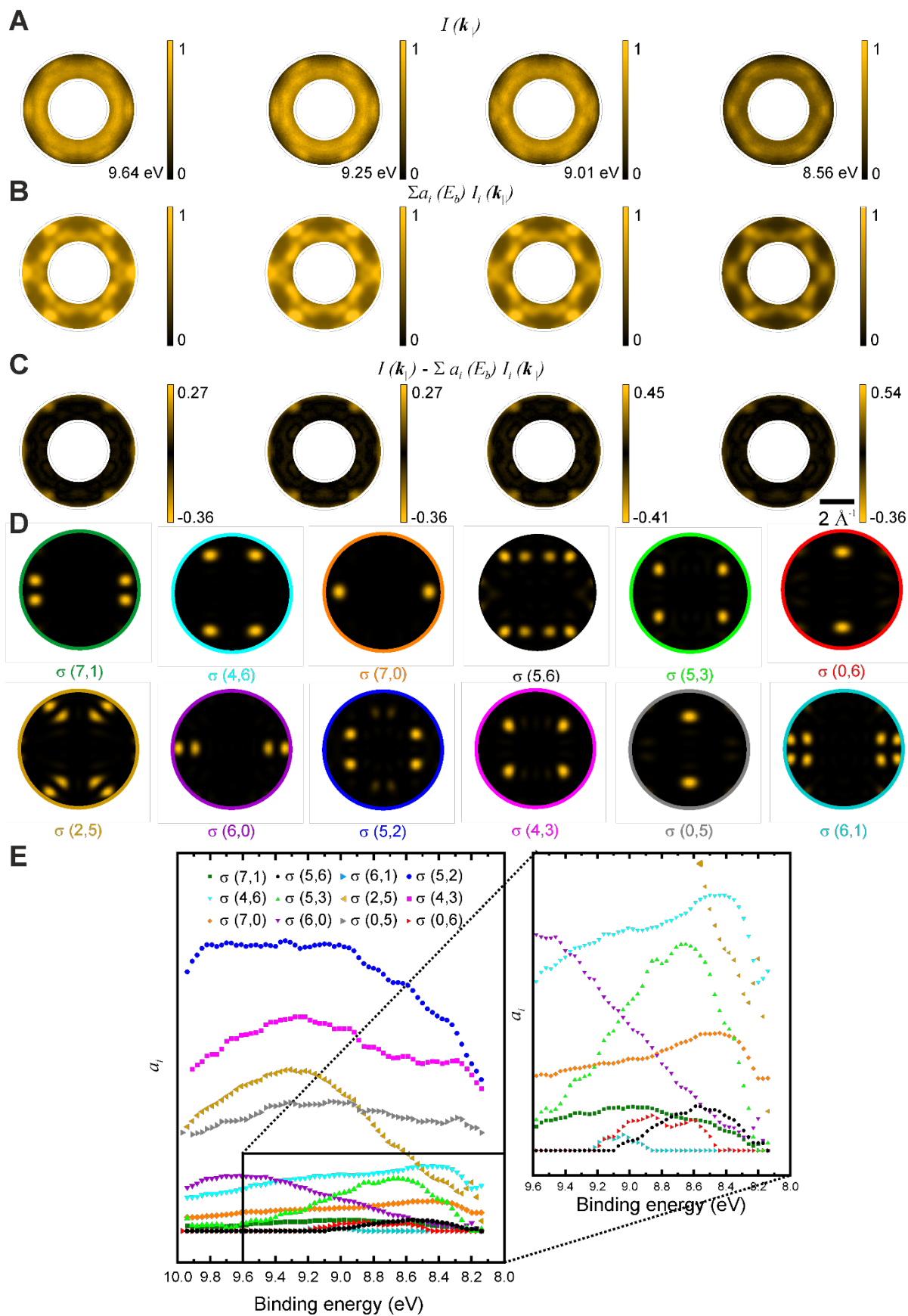

**Fig. S7. Deconvolution of experimental POT results in $E_b$ range of 8.15–10.00 eV.**
(**A**) Experimental $\mathbf{k}_\parallel$-maps $I(\mathbf{k}_\parallel)$ measured at corresponding binding energies. (**B**) $\mathbf{k}_\parallel$-maps $\sum_i a_i(E_b) I_i(\mathbf{k}_\parallel)$ constructed as intensity sum of theoretical $\mathbf{k}_\parallel$-maps $I_i(\mathbf{k}_\parallel)$ of free $C_{28}H_{14}$ shown in **D** and weighted using the fitting parameters $a_i$ from **E**. (**C**) $\mathbf{k}_\parallel$-maps of the deconvolution residual $I(\mathbf{k}_\parallel) - \sum_i a_i(E_b) I_i(\mathbf{k}_\parallel)$. (**D**) Theoretical $\mathbf{k}_\parallel$-maps $I_i(\mathbf{k}_\parallel)$ of free $C_{28}H_{14}$ of the σ-orbitals used for deconvolution. (**E**) Fitting parameters $a_i$ – the experimental pDOS of the σ-orbitals obtained from deconvolution. All $\mathbf{k}_\parallel$-maps are restricted to $|\mathbf{k}_\parallel| \geq 1.7$ Å$^{-1}$ to exclude low $|\mathbf{k}_\parallel|$ emissions from π states.

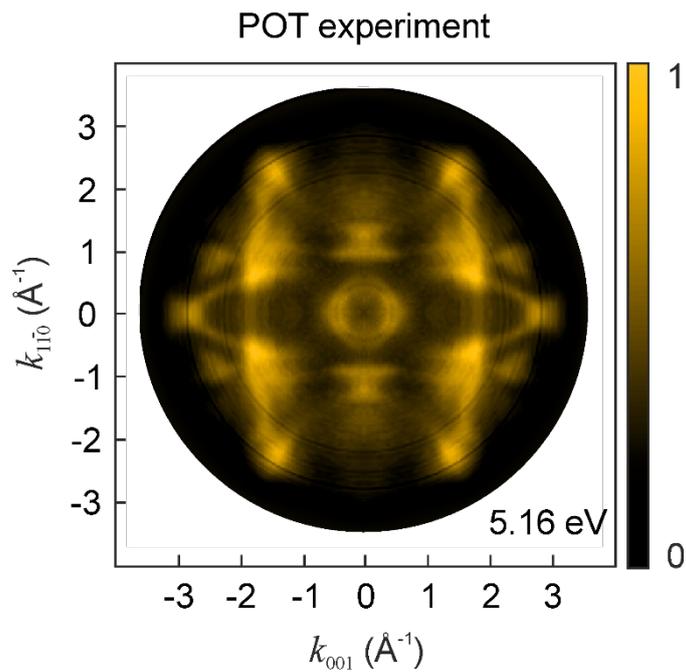

**Fig. S8. Experimental k$_\parallel$-map of clean Cu(110).** Experimental k$_\parallel$-map recorded at $E_b$ =5.16 eV and $h\nu$ =57 eV. A uniform background has been subtracted from the map

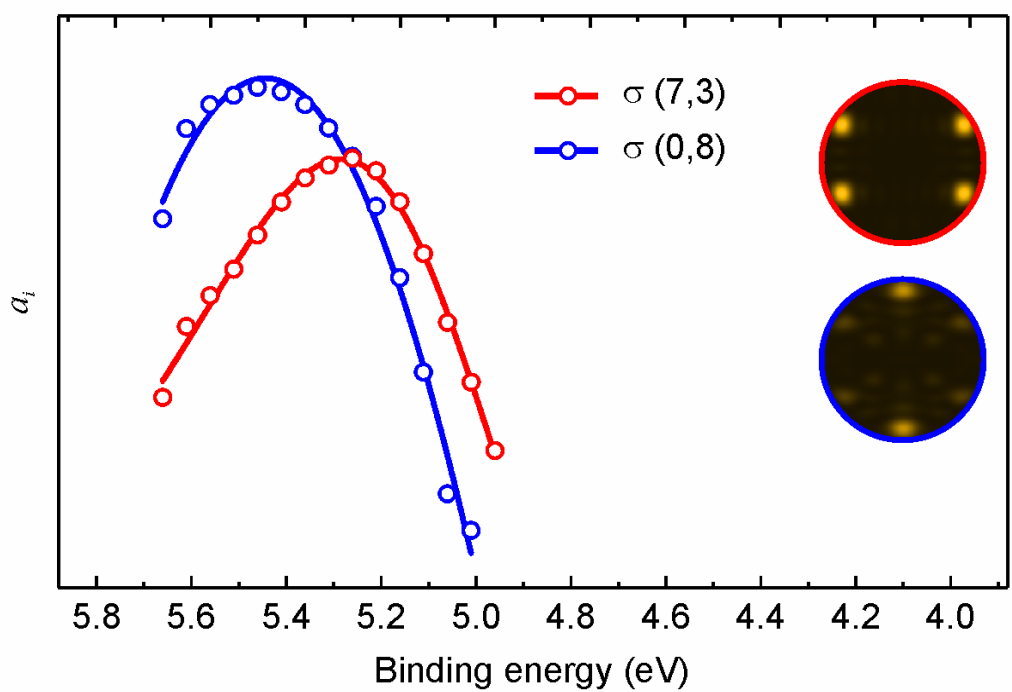

**Fig. S9. Deconvolution of experimental POT results.** Experimental pDOS of σ(7,3) (red) and σ(0,8) (blue) orbitals. The data points were obtained from the deconvolution of a $I(E_{kin}, \mathbf{k}_\parallel)$ data cube, using the theoretical $\mathbf{k}_\parallel$-maps of free $C_{28}H_{12}Cu_2$ shown in the inset. Solid lines are Gaussian fits to the data points.

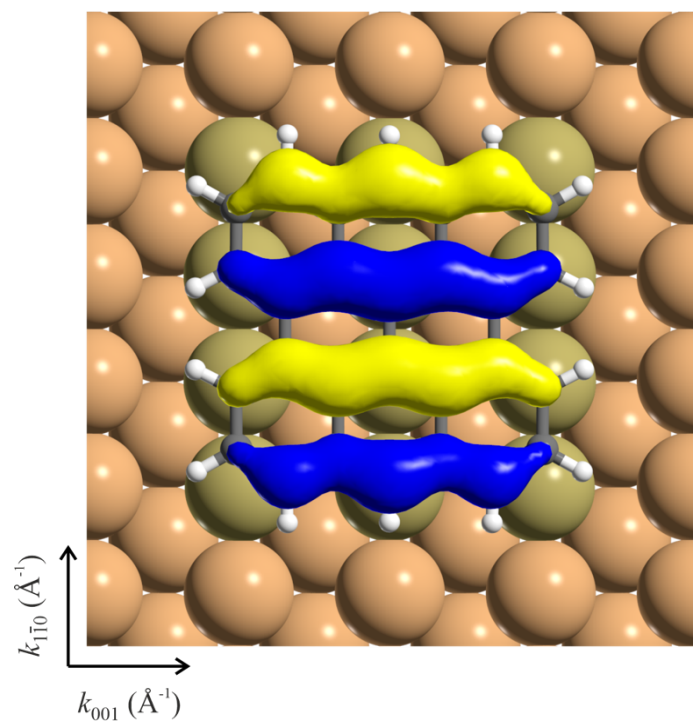

**Fig. S10. Bisanthene on Cu(110).** The energetically favored adsorption site of $C_{28}H_{14}$ on Cu(110) according to DFT. The $\pi(0,3)$ orbital calculated by DFT for free $C_{28}H_{14}$ is superimposed with the molecular backbone.

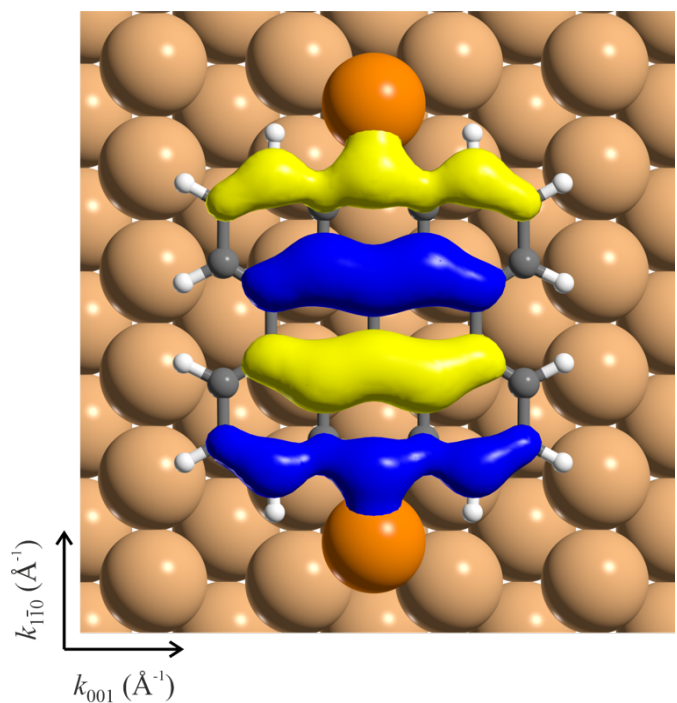

**Fig. S11. Metalated bisanthene on Cu(110).** The energetically favored adsorption site of $C_{28}H_{12}Cu_2$ on Cu(110) according to DFT. The $\pi(0,3)$ orbital calculated by DFT for free $C_{28}H_{12}Cu_2$ is superimposed with the molecular backbone.